\documentclass[12pt,preprint]{aastex}

\shorttitle{Geometric Triangulation to Track CMEs Out to 1 AU}

\shortauthors{Liu et al.}

\begin{document}

\title{Geometric Triangulation of Imaging Observations to
Track Coronal Mass Ejections Continuously Out to 1 AU}

\author{Ying Liu\altaffilmark{1}, Jackie A. Davies\altaffilmark{2},
Janet G. Luhmann\altaffilmark{1}, Angelos Vourlidas\altaffilmark{3},
Stuart D. Bale\altaffilmark{1}, and Robert P. Lin\altaffilmark{1,4}}

\altaffiltext{1}{Space Sciences Laboratory, University of
California, Berkeley, CA 94720, USA; liuxying@ssl.berkeley.edu.}

\altaffiltext{2}{Space Science and Technology Department, Rutherford
Appleton Laboratory, Didcot, UK.}

\altaffiltext{3}{Space Science Division, Naval Research Laboratory,
Washington, DC 20375, USA.}

\altaffiltext{4}{School of Space Research, Kyung Hee University,
Yongin, Gyeonggi 446-701, Korea.}

\begin{abstract}

We describe a geometric triangulation technique, based on
time-elongation maps constructed from imaging observations, to track
coronal mass ejections (CMEs) continuously in the heliosphere and
predict their impact on the Earth. Taking advantage of stereoscopic
imaging observations from STEREO, this technique can determine the
propagation direction and radial distance of CMEs from their birth
in the corona all the way to 1 AU. The efficacy of the method is
demonstrated by its application to the 2008 December 12 CME, which
manifests as a magnetic cloud (MC) from in situ measurements at the
Earth. The predicted arrival time and radial velocity at the Earth
are well confirmed by the in situ observations around the MC. Our
method reveals non-radial motions and velocity changes of the CME
over large distances in the heliosphere. It also associates the
flux-rope structure measured in situ with the dark cavity of the CME
in imaging observations. Implementation of the technique, which is
expected to be a routine possibility in the future, may indicate a
substantial advance in CME studies as well as space weather
forecasting.

\end{abstract}


\section{Introduction}

Coronal mass ejections (CMEs) are large-scale expulsions of plasma
and magnetic field from the solar atmosphere and have been
recognized as primary drivers of interplanetary disturbances.
Tracking CMEs continuously from the Sun to the Earth is crucial for
at least three aspects: a practical capability in space weather
forecasting which has important consequences for life and technology
on the Earth and in space; accurate measurements of CME kinematics
over an extensive region of the heliosphere that are needed to
constrain global MHD simulations of CME evolution; determination of
CME properties from imaging observations that can be properly
compared with in situ data.

CMEs and shocks have been tracked continuously in the heliosphere
using type II radio emissions \citep[e.g.,][]{reiner07, liu08} and
MHD propagation of observed solar wind disturbances
\citep[e.g.,][]{wang01, richardson05, richardson06, liu06a, liu08}.
The frequency drift of type II bursts can be used to characterize
shock propagation but relies on a density model to convert
frequencies to heliocentric distances; in situ measurements of shock
parameters at 1 AU are also needed to constrain the overall
height-time profile due to ambiguities in the frequency drift. MHD
propagation of the solar wind, connecting in situ measurements at
different distances, has been performed only for CME/shock
propagation beyond 1 AU as confined by availability of in situ
measurements close to the Sun. None of these techniques can
determine the propagation direction.

Accurate determination of the propagation direction is feasible with
multiple views of the Sun-Earth space from the Solar Terrestrial
Relations Observatory \citep[STEREO;][]{kaiser08}. Geometric
triangulation techniques using the stereoscopic imaging observations
have been developed to determine CME propagation direction and
radial distance \citep[e.g.,][]{pizzo04, howard08, maloney09}. All
of these methods require the identification and tracking of the same
feature in image pairs from the two spacecraft. This is not possible
at large distances where CME signals become very faint and
diffusive, especially in the field of view (FOV) of the heliospheric
imagers (HI1 and HI2). Construction of time-elongation maps from
imaging observations, which can sense weak signals, has been
extended to HI1 and HI2 data \citep[e.g.,][]{sheeley08, rouillard08,
davies09}. A kinematic model with various assumptions on the
acceleration or velocity of the transient activity is used to fit
the tracks in the time-elongation plot \citep[][]{sheeley99}; in
order to reduce the number of free parameters, CMEs are assumed to
move radially at a constant velocity in the FOV of HI1 and HI2
\citep[e.g.,][]{sheeley08, rouillard08, davies09}. This is a least
squares fit and deals with the minimization of a chi-square
statistic; without apriori knowledge, it is difficult to check
whether the solutions have converged to a global minimum or just a
local one. Therefore, CME kinematics (especially the propagation
direction) cannot be determined unambiguously with the track fitting
technique. In addition, the model fit does not take advantage of
geometric triangulation even though the fit can be performed
independently for the two spacecraft. The full promise of the twin
stereo views has yet to be realized.

In this Letter, we incorporate geometric triangulation in the
time-elongation map analysis using imaging observations from STEREO.
The advantage of this method is that, first, it is based on
time-elongation plots, so geometric triangulation can be applied to
weak features in HI1 and HI2 for the first time; second, it relies
on fewer assumptions than the single track fitting technique, so the
solution is more accurate; third, it can determine the propagation
direction and true distance of CME features (or other white-light
features) from the Sun all the way to 1 AU. This technique is
relatively robust and may indicate an important advance for CME
studies and space weather forecasting.

\section{Instruments and Methodology}

Figure~1 shows the configuration of the two spacecraft with respect
to the Sun. We focus on CME propagation in the ecliptic plane, since
from the perspective of space weather prediction it is important to
know whether/when a CME will impact the Earth. STEREO A is moving
faster and slightly closer to the Sun than the Earth, while STEREO B
is a little further and trailing the Earth; the angular separation
between each spacecraft and the Earth increases by about
22.5$^{\circ}$ per year. Each spacecraft carries an identical
imaging suite, the Sun Earth Connection Coronal and Heliospheric
Investigation \citep[SECCHI;][]{howardra08}, which consists of an
EUV imager (EUVI), two coronagraphs (COR1 and COR2), and two
heliospheric imagers (HI1 and HI2). COR1 and COR2 have an FOV of
0.4$^{\circ}$ - 1$^{\circ}$ and 0.7$^{\circ}$ - 4$^{\circ}$ around
the Sun, respectively. HI1 has a 20$^{\circ}$ square FOV centered at
14$^{\circ}$ elongation from the center of the Sun while HI2 has a
70$^{\circ}$ FOV centered at 53.7$^{\circ}$ from the Sun center.
Combined together these cameras can image a CME from its nascent
stage in the corona all the way to the Earth \citep[see the
accompanying animations online; also see][]{liu09}.

This new instrumentation, especially the HIs, seems so useful that
it will be routinely used in the future. However, geometric
triangulation of HI1 and HI2 data taking advantage of the two
vantage points off the Sun-Earth line has not been implemented since
CME signals in their FOVs are often weak and diffusive. Here we
develop a geometric triangulation technique, applicable to all these
cameras, to determine the radial distance and propagation direction
of CMEs. As shown in Figure~1, a white-light feature would be seen
by the two spacecraft as long as it moves along a direction between
them. The elongation angle of the feature (the angle of the feature
with respect to the Sun-spacecraft line), denoted as $\alpha_A$ and
$\alpha_B$ for STEREO A and B respectively, can be measured from
imaging observations transformed into a Sun-centered coordinate
system. This simple geometry gives
\begin{equation}
\frac{r\sin(\alpha_A+\beta_A)}{\sin\alpha_A} = d_A,
\end{equation}
\begin{equation}
\frac{r\sin(\alpha_B+\beta_B)}{\sin\alpha_B} = d_B,
\end{equation}
\begin{equation}
\beta_A + \beta_B = \gamma,
\end{equation}
where $r$ is the radial distance of the feature from the Sun,
$\beta_A$ and $\beta_B$ are the propagation angles of the feature
relative to the Sun-spacecraft line, $d_A$ and $d_B$ are the
distances of the spacecraft from the Sun (known), and $\gamma$ is
the longitudinal separation between the two spacecraft (also known).
Once the elongation angles ($\alpha_A$ and $\alpha_B$) are measured
from imaging observations, the above equations can be solved for
$r$, $\beta_A$ and $\beta_B$, a unique set of solutions (compared
with model fit). For $d_A = d_B$, these equations can be reduced to
\begin{equation}
\tan\beta_A = \frac{\sin\alpha_A\sin(\alpha_B+\gamma) -
\sin\alpha_A\sin\alpha_B}{\sin\alpha_A\cos(\alpha_B+\gamma) +
\cos\alpha_A\sin\alpha_B},
\end{equation}
which is generally true for STEREO A and B and allows a quick
estimate of the propagation direction.

The elongation angles can be obtained from time-elongation plots
produced by stacking the running difference intensities along the
ecliptic plane. Even weak signals are discernible in these maps, so
transient activity can be revealed over an extensive region of the
heliosphere. CME features usually appear as tracks extending to
large elongation angles in the maps. Previous studies use only
equation~(1) or (2) to fit the tracks assuming a kinematic model
with a constant propagation direction \citep[e.g.,][]{sheeley99,
sheeley08, rouillard08, davies09, davis09}.

Our only assumption is that the same feature can be tracked in both
the time-elongation maps from the two spacecraft. This is likely
true if the tracks between the maps have a good timing (see details
below). COR1 and COR2 have a cadence of 5 and 15 minutes,
respectively; tracking of the same feature from the timing is good
to these time scales. However, the imaging observations provide
integrated line-of-sight information through a 3D structure.
Projection and Thomson-scattering effects may affect the tracks in
the time-elongation maps in ways that are difficult to assess
quantitatively without detailed modeling of the coronal brightness
\citep[][]{vourlidas06, lugaz08, lugaz09}. Such effects are
minimized for features propagating symmetrically relative to the two
spacecraft (i.e., along the Sun-Earth line). The uncertainties
brought about by these effects will be addressed by global MHD
simulations in an ongoing work.

\section{Application}

To prove the efficacy of the method, we apply it to the 2008
December 12 CME. Figure~2 shows two synoptic views of the CME from
STEREO A and B. During the time of the CME,
$\gamma\simeq86.3^{\circ}$, $d_A \simeq 0.97$ AU, and $d_B \simeq
1.04$ AU; STEREO A and B are generally within 5$^{\circ}$ of the
ecliptic plane. The latitudinal separation between the two
spacecraft is ignored in our triangulation analysis for simplicity.
A panoramic view of the CME evolution from the low corona to the
Earth is also provided in the online journal as animations made of
composite images from the complete imaging system. The CME is
induced by a prominence eruption in the northern hemisphere, which
started between 03:00 - 04:00 UT on December 12. The prominence
material (visible in EUVI at 304 \AA) is well aligned with the CME
core. The CME slowly rotates and expands toward the ecliptic plane,
and seems fully developed in COR2. Running difference images are
used for HI1 and HI2 to remove the F corona (produced by dust
scattering of the white light) and stellar background. The Earth is
visible in HI2 from both STEREO A and B, while Venus is seen only
from STEREO A; their brightness saturates the detector, resulting in
a vertical line in the image. Oscillation of stars is also visible
in HI2 from STEREO B, probably due to a slight shaking of the
camera. The basic structure of the CME remains organized out to the
FOV of HI1 (close to the Sunward edge), but only wave-like
structures are seen in HI2. Apparently the weak diffusive signal is
difficult to track using traditional techniques.

The time-elongation maps shown in Figure~3 are produced by stacking
running difference intensities of COR2, HI1 and HI2 within a slit of
64 pixels around the ecliptic plane. In the FOVs of HI1 and HI2, the
coronal intensity is dominated by the F corona and stellar
background. The contribution of the F corona is minimized by
subtracting a long-term background from each image. Adjacent images
are aligned prior to the running differencing to remove stars from
the FOV. Both the image alignment and determination of elongation
angles require precise pointing information of the HI cameras, so we
use the level 1 data (available at
\url{http://www.ukssdc.rl.ac.uk/solar/stereo/data.html}) that have
been corrected for flat field, shutterless readout and instrument
offsets from the spacecraft pointing. We apply a median filter to
these difference images to reduce the residual stellar effects. A
radial strip with a width 64 pixels around the ecliptic plane is
extracted from each difference image, and a resistant mean is taken
over the 64 pixels to represent the intensity at corresponding
elongation angles. These resistant means are then stacked as a
function of time and elongation, and the resulting map is scaled to
enhance transient signals.

In Figure~3, two features coincident with the CME can be identified
up to 50$^{\circ}$ elongation for both STEREO A and B. For
comparison, the Earth is at an elongation angle 70$^{\circ}$ for
STEREO A and 64$^{\circ}$ for STEREO B. The temporal coincidence of
each track between the two maps indicates that we are tracking the
same feature. Intermittent ones between the two tracks, probably
associated with the CME core, are also seen but later disappear
presumably due to the expansion of the CME (see Figure~5). Note that
the elongation angles are plotted in a logarithmic scale to expand
COR2 data; tracks are not J-like as in traditional linear-linear
plots. Elongation angles are extracted along the trailing edge of
these two tracks (with the sharpest contrast); interpolation is then
performed to get elongation angles at the same time tags for STEREO
A and B as required by the triangulation analysis. The uncertainty
in the measurements of elongation angles is estimated to be about 10
pixels, which is roughly 0.02$^{\circ}$, 0.2$^{\circ}$ and
0.7$^{\circ}$ for COR2, HI1 and HI2 respectively. We input the
values of the elongation angles to equations (1)-(4) to calculate
the propagation direction and radial distance. Time-elongation maps
from COR1 images are also examined but not included here, given its
small FOV and the fact that the CME is largely above the ecliptic
plane through COR1.

The resulting CME kinematics are displayed in Figure~4. The
propagation direction ($\beta_A$ or $\beta_B$) is converted to an
angle with respect to the Sun-Earth line. If the angle is positive
(negative), the CME feature would be propagating west (east) of the
Sun-Earth line in the ecliptic plane. The propagation direction (for
both features 1 and 2) shows a variation with distance but is
generally within 5$^{\circ}$ of the Sun-Earth line. These features
can be continuously tracked up to 150 solar radii or 0.7 AU (without
projection). Radial velocities are derived from the distance using a
numerical differentiation with three-point Lagrangian interpolation.
The radial velocity also shows a variation with distance: it first
increases rapidly and then decreases (clearer for feature 1 which is
the CME leading edge). The radial velocity is about $363\pm 43$ km
s$^{-1}$ for feature 1 and $326\pm 51$ km s$^{-1}$ for feature 2
close to the Earth, estimated by averaging data points after
December 14. Note that, different from previous studies, our method
is unique since it can determine CME kinematics (both propagation
direction and radial velocity) as a function of distance from the
Sun all the way to 1 AU; the CME kinematics determined this way
provide an unprecedented opportunity to constrain global MHD models
of CME evolution.

We test these results using in situ measurements. Figure~5 shows a
magnetic cloud (MC) identified from WIND data based on the strong
magnetic field and smooth rotation of the field. A similar field and
velocity structure is also observed at ACE, but ACE does not have
valid measurements of proton density and temperature due to the low
solar wind speeds during this time interval (R. M. Skoug 2009,
private communication). The MC radial width (average speed times the
duration) is about 0.1 AU, relatively small compared with the
average level 0.2 - 0.3 AU at 1 AU \citep[e.g.,][]{liu05, liu06b}.
It is remarkable that even such a small event can be tracked to
those large distances. The MC density is lower than that of the
ambient solar wind, presumably owing to the expansion as shown by
the declining speed profile across the MC. The MC has a
well-organized magnetic field structure; our reconstruction with the
Grad-Shafranov method gives a left-handed flux-rope configuration.
This is surprising given the imprint of a rather diffusive
morphology in HI2 images. The predicted arrival times of features 1
and 2, obtained with a second-order polynomial fit of the radial
distance shown in Figure~4, bracket the MC and are coincident with
density-enhanced structures around the MC. It is interesting that
these high-density structures, which appear to be part of the CME in
imaging observations, are actually not contained in the flux rope.
This finding supports the view that the flux rope corresponds to the
dark cavity in the classic three-part structure of CMEs (front,
cavity and core). The arrival time prediction is good to within a
few hours. Note that we are not tracking the flux rope but
density-enhanced regions before and after the MC, and the CME front
has swept up and merged with the ambient solar wind during its
propagation in the heliosphere. The predicted radial velocities at 1
AU, an average over the data points after December 14 in Figure~4,
are also well confirmed by in situ measurements.

The good agreement between the geometric triangulation analysis and
in situ data demonstrates a technique that can predict CME impact
well in advance; remote sensing observations can also be properly
connected with in situ measurements with the aid of the method.
Various attempts have also been made to predict CME arrival times at
the Earth with an accuracy ranging from 2 - 11 hours
\citep[e.g.,][]{zhao02, michalek04, howard06}. The present method is
expected to give a more accurate prediction since it can track CMEs
from the Sun continuously to 1 AU. A statistical study, however, is
needed before we affirm the effectiveness of the technique.

\citet{davis09} studied the same event based on a single spacecraft
analysis (HI data only). They assumed a kinematic model with a
constant velocity and propagation direction, which has the least
free parameters. Their results are roughly similar to ours, except
that they obtain a radial velocity larger than observed and a larger
uncertainty in the propagation direction (see their Table 1). It is
not surprising that, given such a small CME, their assumptions of
constant velocity and propagation direction in the FOV of the HIs
are more or less verified by our results. These assumptions are
usually not true for CME propagation close to the Sun (see
Figure~4). Fast CMEs would have a strong interaction with the
heliosphere and thus their velocities vary with distance even far
away from the Sun \citep[e.g.,][]{liu08}. The single track fitting
approach, however, still remains useful especially when data are
only available from one spacecraft.

\section{Summary}

We have presented a geometric triangulation technique to track CMEs
from the Sun all the way to 1 AU, based on stereoscopic imaging
observations of STEREO. This new method enables geometric
triangulation for HI1 and HI2 data, relaxes various assumptions
needed in a model fit \citep[e.g.,][]{sheeley99, sheeley08,
rouillard08, davies09}, and can accurately determine CME kinematics
(both propagation direction and radial distance) continuously in the
heliosphere. A good agreement between the geometric triangulation
analysis and in situ measurements is obtained when we apply the
method to the 2008 December 12 CME. Velocity changes and non-radial
motions of the CME are also revealed by the tracking technique; this
propagation history over a large distance is crucial to probe CME
interaction with the heliosphere. This method enables a better
connection between imaging and in situ observations. It fulfills a
major objective of the STEREO mission, and also heralds a new era
when CMEs can be tracked over a far more extensive region of the
heliosphere than previously possible with coronagraph observations.

\acknowledgments The research was supported by the STEREO project
under grant NAS5-03131. SECCHI was developed by a consortium of NRL,
LMSAL and GSFC (US), RAL and Univ. Birmingham (UK), MPI (Germany),
CSL (Belgium), and IOTA and IAS (France). We acknowledge the use of
WIND data. R. Lin has been supported in part by the WCU grant (No.
R31-10016) funded by KMEST.

\clearpage

\begin{figure}
\epsscale{0.8} \plotone{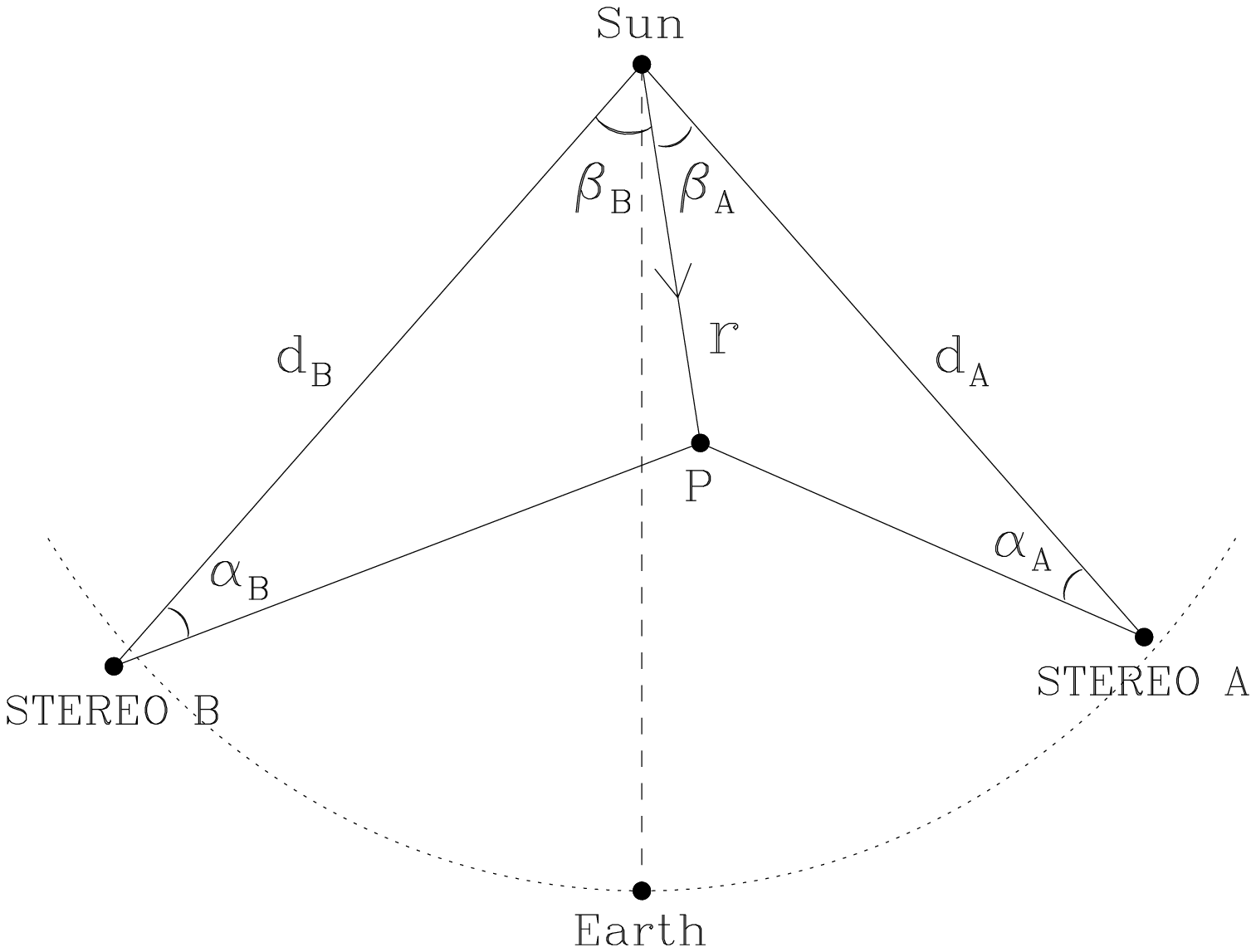} \caption{Diagram for the geometric
triangulation in the ecliptic plane. The dotted line indicates the
orbit of the Earth while the dashed line represents the Sun-Earth
line. A white-light feature, propagating along the direction shown
by the arrow, is denoted by the point P.}
\end{figure}

\clearpage

\begin{figure}
\epsscale{0.56} \plotone{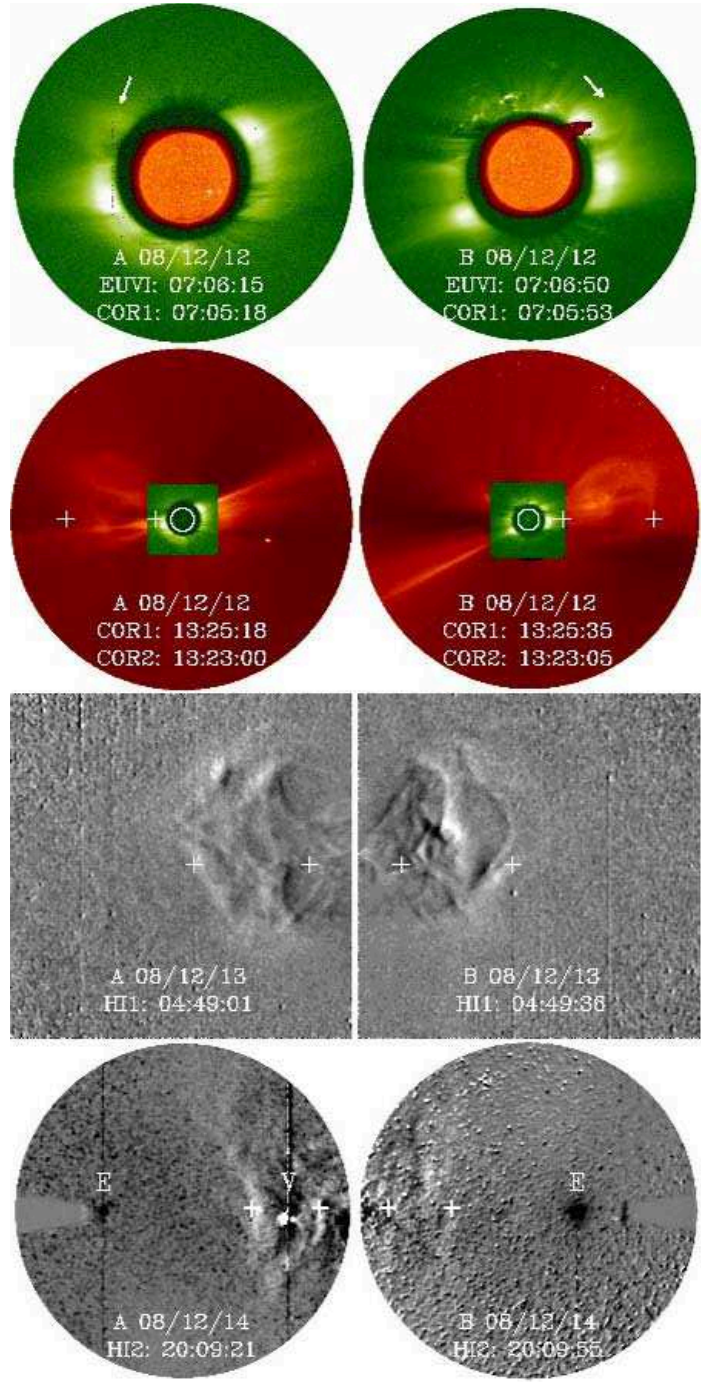} \caption{CME evolution observed by
STEREO A (left) and B (right) near simultaneously. From top to
bottom, the panels display the composite images of EUVI at 304 \AA\
and COR1 showing the nascent CME (indicated by the arrow), combined
COR1 and COR2 images of the fully developed CME, and running
difference images from HI1 and HI2 when the CME is far away from the
Sun. The crosses mark the locations of the two features obtained
from Figure~3. The positions of the Earth and Venus are labeled as E
and V. (Mpeg animations made of composite images are available in
the online journal.)}
\end{figure}

\clearpage

\begin{figure}
\epsscale{0.7} \plotone{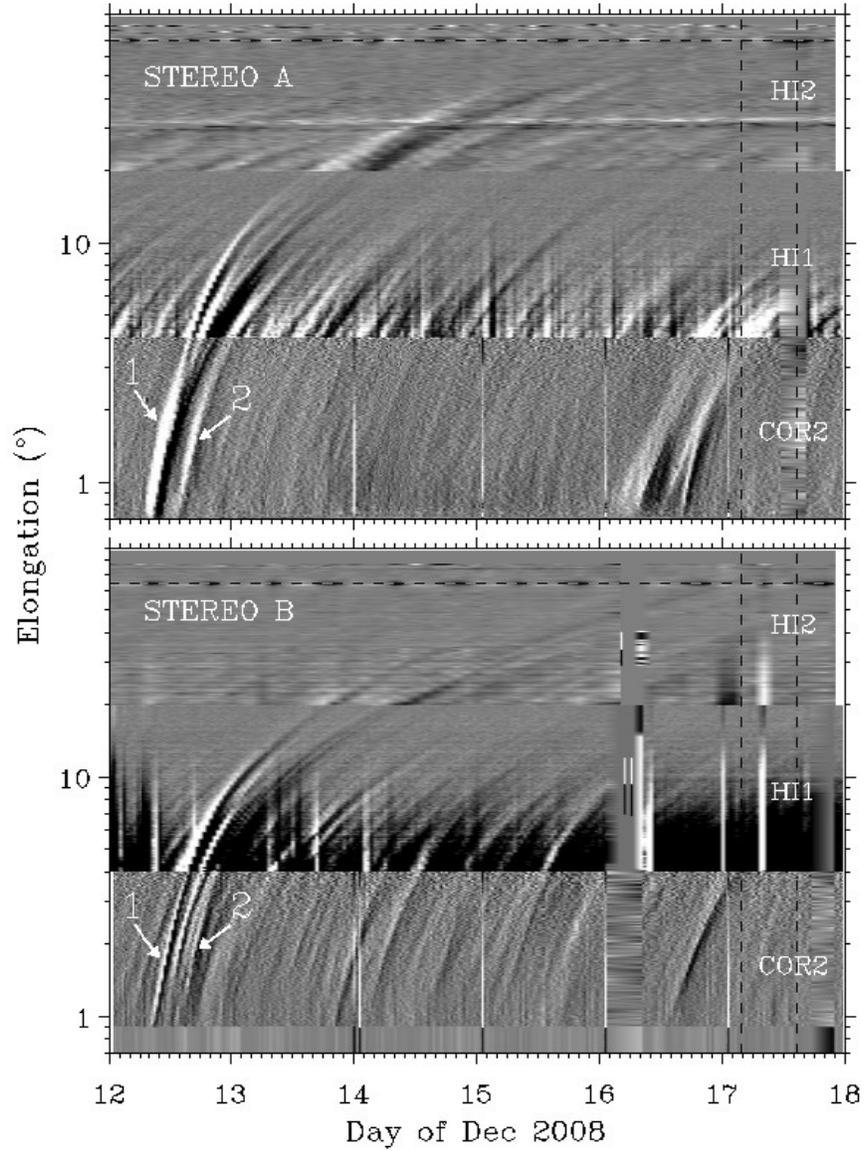} \caption{Time-elongation maps
constructed from running difference images of COR2, HI1 and HI2
along the ecliptic plane for STEREO A (upper) and B (lower). The
arrows indicate two tracks associated with the CME. The vertical
dashed lines show the MC interval observed at WIND, and the
horizontal dashed line marks the elongation angle of the Earth.}
\end{figure}

\clearpage

\begin{figure}
\epsscale{0.7} \plotone{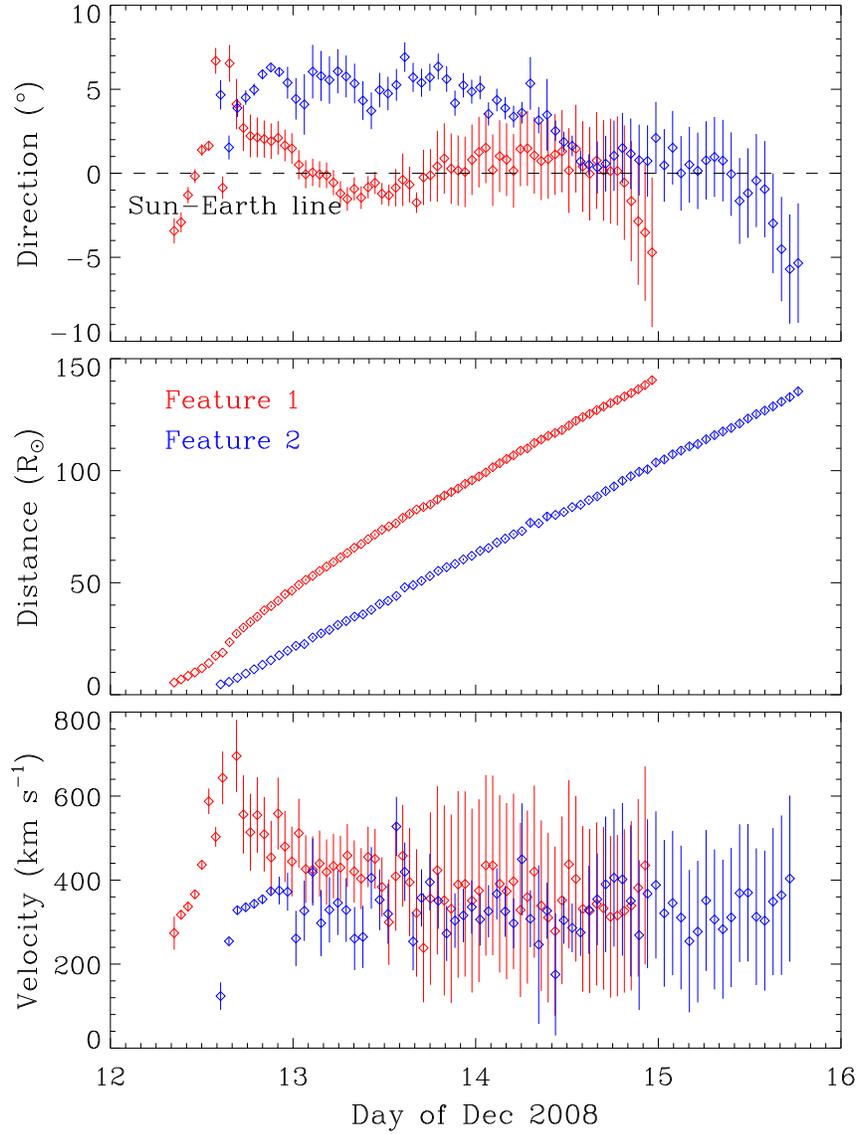} \caption{Propagation direction,
radial distance and velocity of features 1 (red) and 2 (blue)
derived from the geometric triangulation analysis. The dashed line
indicates the Sun-Earth line. Error bars represent uncertainties
mathematically derived from the measurements of elongation angles.
Note that the velocities are calculated from adjacent distances and
often have misleading error bars.}
\end{figure}

\clearpage

\begin{figure}
\epsscale{0.7} \plotone{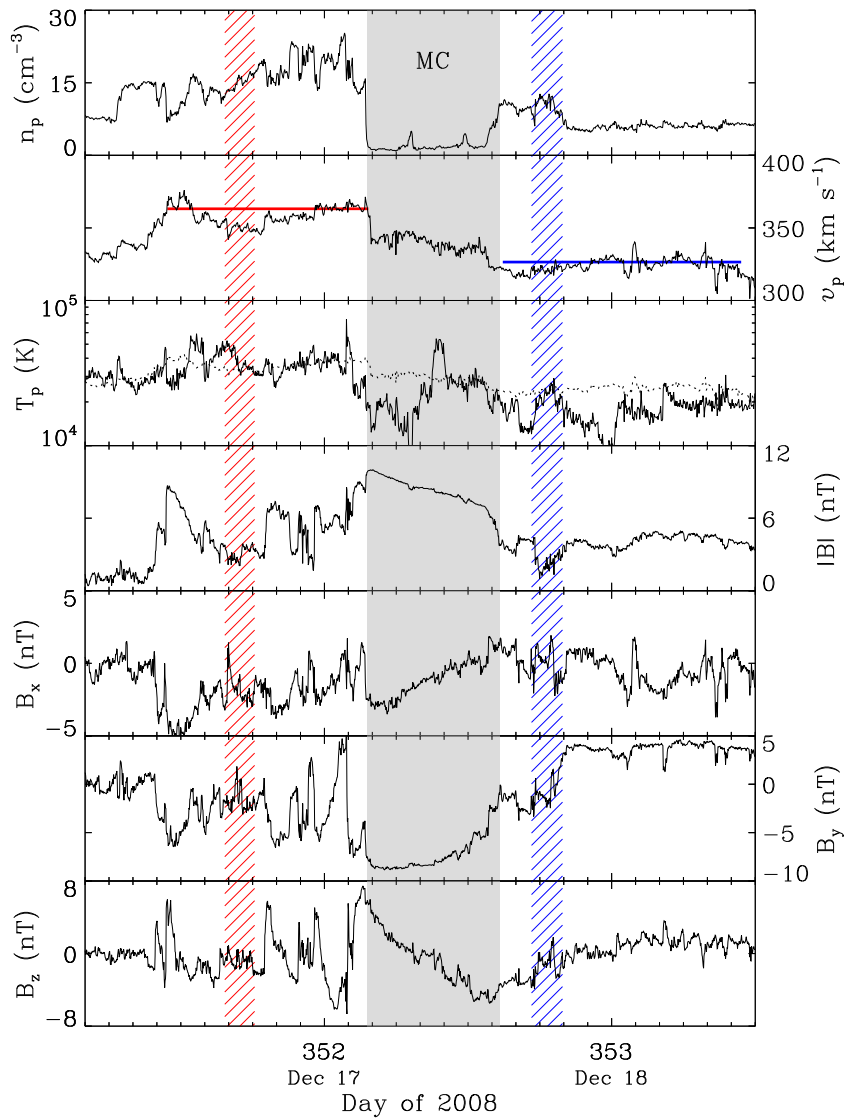} \caption{Solar wind plasma and
magnetic field parameters across the MC observed at WIND. From top
to bottom, the panels show the proton density, bulk speed, proton
temperature, and magnetic field strength and components,
respectively. The shaded region indicates the MC interval, and the
hatched area shows the predicted arrival times (with uncertainties)
of features 1 (red) and 2 (blue). The horizontal lines mark the
corresponding predicted velocities at 1 AU. The dotted line denotes
the expected proton temperature from the observed speed.}
\end{figure}

\end{document}